\documentclass[aps,pra,reprint,superscriptaddress,showpacs]{revtex4-1}

\usepackage{graphicx}
\usepackage{dcolumn}
\usepackage{bm}
\usepackage{hyperref}
\usepackage{amsmath,amssymb}
\usepackage{physics}
\DeclareMathOperator{\sinc}{sinc}
\usepackage{mathrsfs}
\usepackage{mhchem}
\usepackage{siunitx}
\usepackage{soul, xcolor}

\setstcolor{violet}

\begin{document}


\title{Idealised EPR states from non-phase matched parametric down conversion}

\author{C. Okoth}
\affiliation{Max Planck Institute for the Science of Light, Staudtstra{\ss}e 2, 91058 Erlangen, Germany}
\affiliation{University of Erlangen-N\"urnberg, Staudtstra{\ss}e 7/B2, 91058 Erlangen, Germany}
\author{E. Kovlakov}
\affiliation{Department of Physics, M. V. Lomonosov Moscow State University, Leninskie Gory, 119991 Moscow, Russia}
\affiliation{Quantum Technologies Centre, M. V. Lomonosov Moscow State University, Leninskie Gory, 119991 Moscow, Russia}
\author{F. B{\"o}nsel}
\affiliation{Max Planck Institute for the Science of Light, Staudtstra{\ss}e 2, 91058 Erlangen, Germany}
\affiliation{University of Erlangen-N\"urnberg, Staudtstra{\ss}e 7/B2, 91058 Erlangen, Germany}
\author{A. Cavanna}
\affiliation{Max Planck Institute for the Science of Light, Staudtstra{\ss}e 2, 91058 Erlangen, Germany}
\affiliation{University of Erlangen-N\"urnberg, Staudtstra{\ss}e 7/B2, 91058 Erlangen, Germany}
\author{S. Straupe}
\affiliation{Department of Physics, M. V. Lomonosov Moscow State University, Leninskie Gory, 119991 Moscow, Russia}
\affiliation{Quantum Technologies Centre, M. V. Lomonosov Moscow State University, Leninskie Gory, 119991 Moscow, Russia}
\author{S. P. Kulik}
\affiliation{Department of Physics, M. V. Lomonosov Moscow State University, Leninskie Gory, 119991 Moscow, Russia}
\affiliation{Quantum Technologies Centre, M. V. Lomonosov Moscow State University, Leninskie Gory, 119991 Moscow, Russia}
\author{M. V. Chekhova}
\affiliation{Max Planck Institute for the Science of Light, Staudtstra{\ss}e 2, 91058 Erlangen, Germany}
\affiliation{University of Erlangen-N\"urnberg, Staudtstra{\ss}e 7/B2, 91058 Erlangen, Germany}
\affiliation{Department of Physics, M. V. Lomonosov Moscow State University, Leninskie Gory, 119991 Moscow, Russia}

\date{\today}
             
\begin{abstract}

Entanglement of high dimensional states is becoming increasingly important for quantum communication and computing. The most common source of entangled photons is spontaneous parametric down conversion (SPDC), where the degree of frequency and momentum entanglement is determined by the non-linear interaction volume. Here we show that by reducing the length of a highly non-linear material to the micrometer scale it is possible to relax the longitudinal phase matching condition and reach record levels of transverse wavevector entanglement. From a micro-sized layer of lithium niobate we estimate the number of entangled angular modes to be over 1200. The entanglement is measured both directly using correlation measurements and indirectly using stimulated emission tomography. The high entanglement of the state generated can be used to massively increase the quantum information capacity of photons, but it also opens up the possibility to improve the resolution of many quantum imaging techniques.

\end{abstract}

\maketitle

Entanglement is a unique phenomenon that underpins many quantum technologies and applications, such as quantum imaging \cite{kolobov2007quantum}, quantum communication \cite{gisin2007quantum} and quantum computation \cite{nielsen2002quantum}. Spontaneous parametric down conversion (SPDC), the decay of a single high-energy photon into two lower-energy daughter photons, signal and idler, is a convenient source of entangled photons. It has been used to successfully demonstrate effects such as the violation of Bell's inequalities and quantum teleportation \cite{boschi1998experimental,shih1988new}. Many of these famous experiments have been carried out by measuring the entangled state in a discrete variable basis, for example polarisation. However there has been a growing interest in continuous variables (CV) entanglement, for example in: momentum, frequency and quadrature amplitudes \cite{braunstein2005quantum}. Measurements made in a CV basis provide access to a high-dimensional Hilbert space. This, in turn, allows a large amount of information to be encoded in a relatively small number of photons, making CV entanglement desirable for quantum communication \cite{adachi2007simple,jennewein2000quantum,mair2001entanglement}.

At a fixed frequency and fixed azimuthal angle the two-photon state generated via SPDC in (polar) angular space can be modelled~\cite{monken1998transfer} as 
\begin{equation}
    \ket{\psi_{\theta_i,\theta_s}}=C\int d\theta_i d\theta_s F (\theta_i,\theta_s) a_i^{\dagger}(\theta_i) a_s^{\dagger}(\theta_s)\ket{0_s,0_i}
\end{equation}
where $C$ is a normalization constant. The signal, $a_i^{\dagger}(\theta_i)$, and idler, $a_s^{\dagger}(\theta_s)$, creation operators generate photons into modes with wavevectors that subtend angles $\theta_i$ and $\theta_s$ with respect to the pump direction. The complex amplitude $F (\theta_i,\theta_s)$ dictates the degree of entanglement. In general, $F (\theta_i,\theta_s)$ can be separated into two factors, the pump and phase matching functions, which depend on the transverse wavevector mismatch, $\Delta k_{\bot}(\theta_i,\theta_s)$, and longitudinal wavevector mismatch, $\Delta k_{\parallel}(\theta_i,\theta_s)$, respectively:  $F (\theta_i,\theta_s) = F_{p} (\Delta k_{\bot}) F_{pm} (\Delta k_{\parallel})$. Their widths are determined by the inverse pump beam waist $\sigma$ and inverse length of the non-linear material $L$. The joint probability density of the state, also known as the two-photon intensity (TPI), is given by $\abs{F (\theta_i,\theta_s)}^2$. One can assign an unconditional or marginal distribution to the TPI which gives the single photon angular emission width, $\Delta_{\theta}$, and a conditional distribution which gives the coincidence angular width, $\delta_{\theta}$ \cite{fedorov2008spontaneous,Fedorov2007}. The ratio of the two widths is an operational measure of the degree of entanglement \cite{fedorov2004packet}. 

In the CV basis an idealised EPR state $\ket{\text{EPR}_{\theta_i,\theta_s}}=\int d\theta_i d\theta_s \delta(\theta_i-\theta_s) a_i^{\dagger}(\theta_i) a_s^{\dagger}(\theta_s)\ket{0_s,0_i}$, describes a maximally entangled state \cite{einstein1935can,howell2004realization}. This corresponds to a situation where one photon out of the pair can be emitted into any angle, but once it has been detected the emission angle of the conjugate photon in the pair is known to an infinite degree of accuracy. Therefore if $F (\theta_i,\theta_s)\rightarrow \delta(\theta_i-\theta_s)$, then the two-photon state is maximally entangled. For SPDC this situation is realised when $L$ is decreased to the point that the phase matching function becomes so broad that it can be approximated by 1 everywhere and $\sigma$ is increased to the point that the pump function is so narrow that it can be approximated by a delta function.

\begin{figure}
\includegraphics[width=8.5cm]{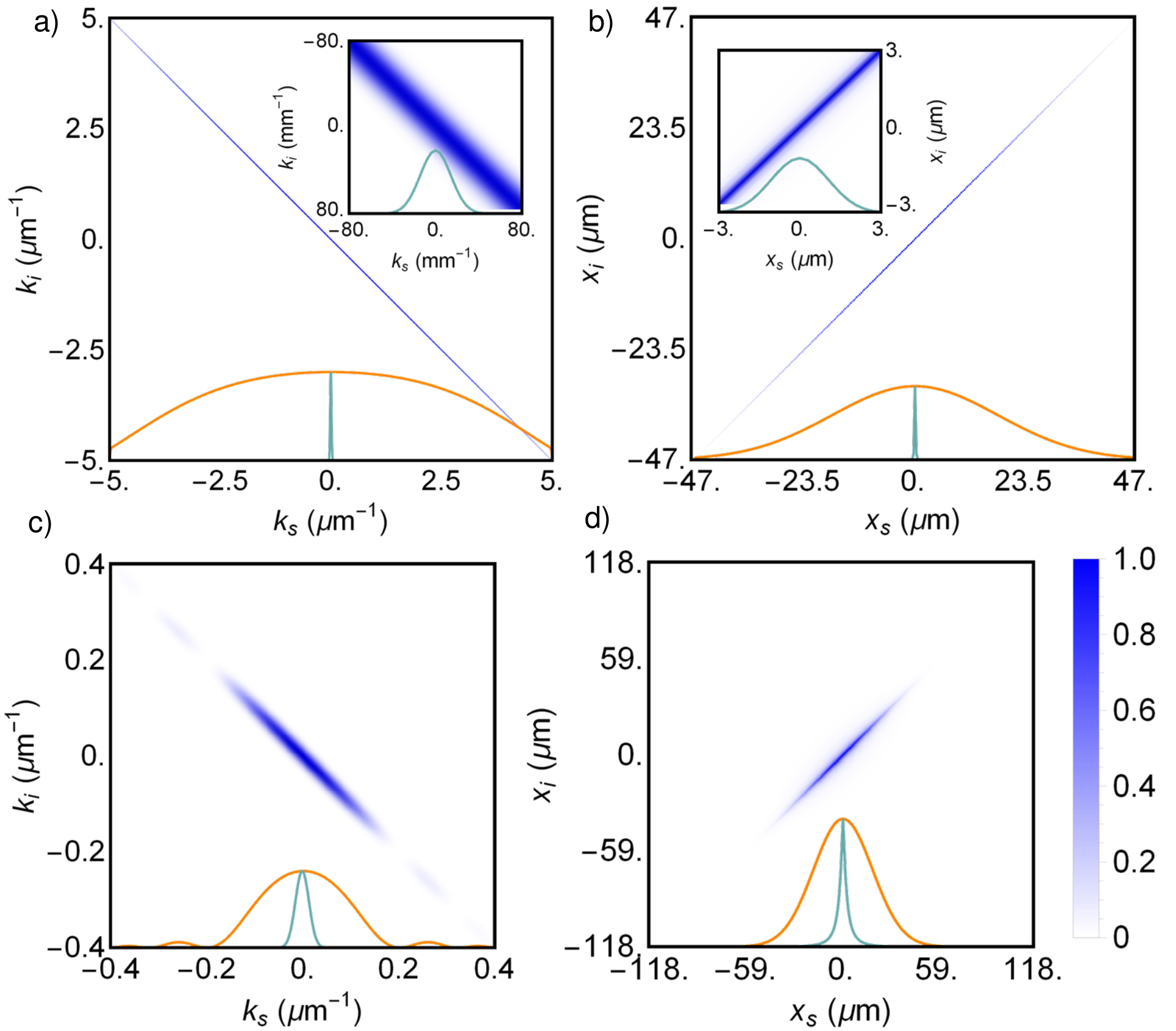} 
\caption{Top: theoretical TPI for non-phase matched type-0 SPDC in LN in the far field (a) and near field (b). The interaction length is $L=1.38 \mu\text{m}$ and the beam waist is $\sigma= 50 \mu\text{m}$. Bottom: TPI for phase-matched type-I collinear SPDC in LN in the far field c) and near field d). The beam waist is again $\sigma= 50 \mu\text{m}$, however the length is $L=1.4 \text{mm}$. The unconditional distributions are given by the orange curves and the conditional distributions are given by the teal curves. The insets in a) and b) show a zoom of the TPI.}
\label{fig:theory}
\end{figure}

Recently two-photon radiation was generated from a microscale length of lithium niobate (LN) \cite{okoth2019non}, which currently is the shortest $L$ reported. At such length scales the phase matching function, $F_{pm} (\Delta k_{\parallel})$, becomes massively broadened leading to the generation of a highly entangled state. Until now SPDC has only been observed with the wavevector mismatch zero or close to zero~\cite{pires2009direct}, such that momentum conservation was satisfied. However, for very small $L$ this condition does not need to be upheld strictly. This opens up the possibility of using materials with large second-order susceptibilities, that are normally disregarded in the phase matched regime. In this way it is possible to partially compensate for the lower SPDC efficiency due to the reduced interaction length by using highly non-linear materials. This leads to the surprisingly efficient generation of a biphoton state via non-phase matched SPDC. The TPI of a non-phase matched biphoton state compared to a phase matched state is shown in Fig.~\ref{fig:theory}.

Whilst the above considers the degree of entanglement in transverse wavevector, represented here as the emission angle, it is also interesting to consider the degree of entanglement in the transverse position. Position (or near-field) entanglement should have the same high value as the momentum (far-field) entanglement~\cite{just2013transverse}. The coincidence width in the near field, $\delta_{x}$, is proportional to the length $L$, whilst the single photon positional width, $\Delta_{x}$, is equal to the pump beam waist $\sigma$. 
If the length $L$ is sufficiently small, then the correlation width in position can be deeply subwavelength whilst the uncertainty in the single photon position is limited by the pump beam waist (see Fig.~\ref{fig:theory}b).          

\begin{figure}
\includegraphics[width=9cm]{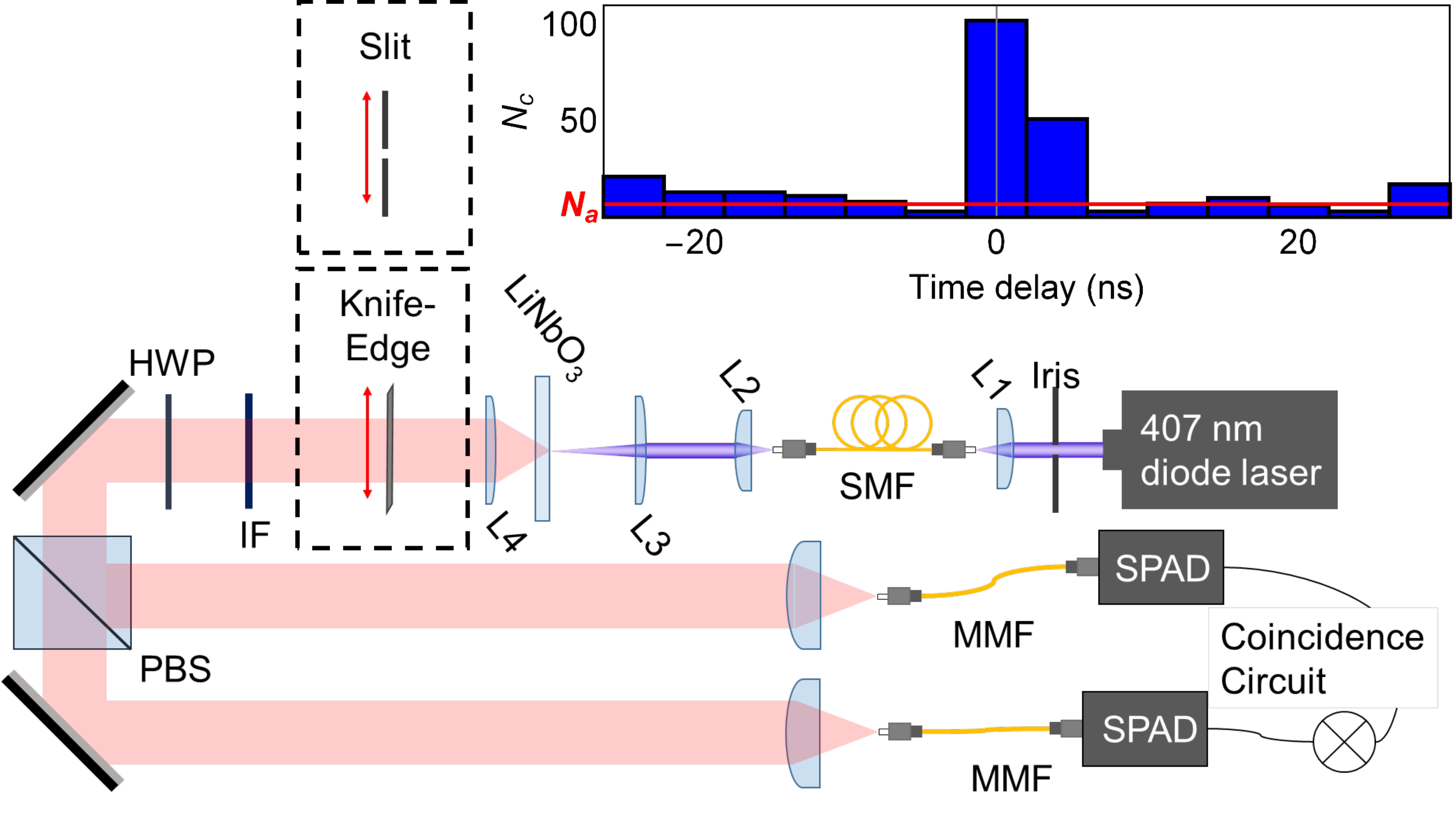} 
\caption{Experimental setup for characterising the TPI from spontaneously emitted photon pairs. The inset shows the number of coincidences $N_c$ for different time delay bins. $N_a$ is the number of accidental coincidences.}
\label{fig:setup1}
\end{figure}

To find the degree of entanglement, both the full angular range $\Delta_{\theta}$ of the photon pair emission and the angular correlation width $\delta_{\theta}$ were measured. This was done using the setup shown in Fig. \ref{fig:setup1}. The pump was a 30 mW continuous wave laser. To ensure that the pump had a good mode quality it was sent through a single-mode fiber (SMF). The pump was focused (L3) into layer of MgO-doped lithium niobate. The 3'' LN wafer was fabricated with an inhomogenous thickness. This meant that by moving the LN in the transverse plane it was possible to tune the interaction length $L$ from 5.8 microns to 6.8 microns. The X-cut LN was oriented such that the polarised pump interacted with the highest component ($\chi^{(2)}_{zzz}$) of the second-order susceptibility. L3 was selected to maximise the degree of entanglement, yet still allow the capture and detection of photon pairs. The chosen optimal focal length was $11$ mm (see Supplementary). Although this led to a relatively low degree of entanglement, the aim was to measure the large unconditional width, which is almost independent of the beam waist size. The resulting emission was collimated using a high NA aspheric lens with a $7.5$ mm focal length (L4), to ensure that the full angular range of the emission was being collected, and the pump radiation was filtered (IF) out.  

\begin{figure}
\includegraphics[width=8.5cm]{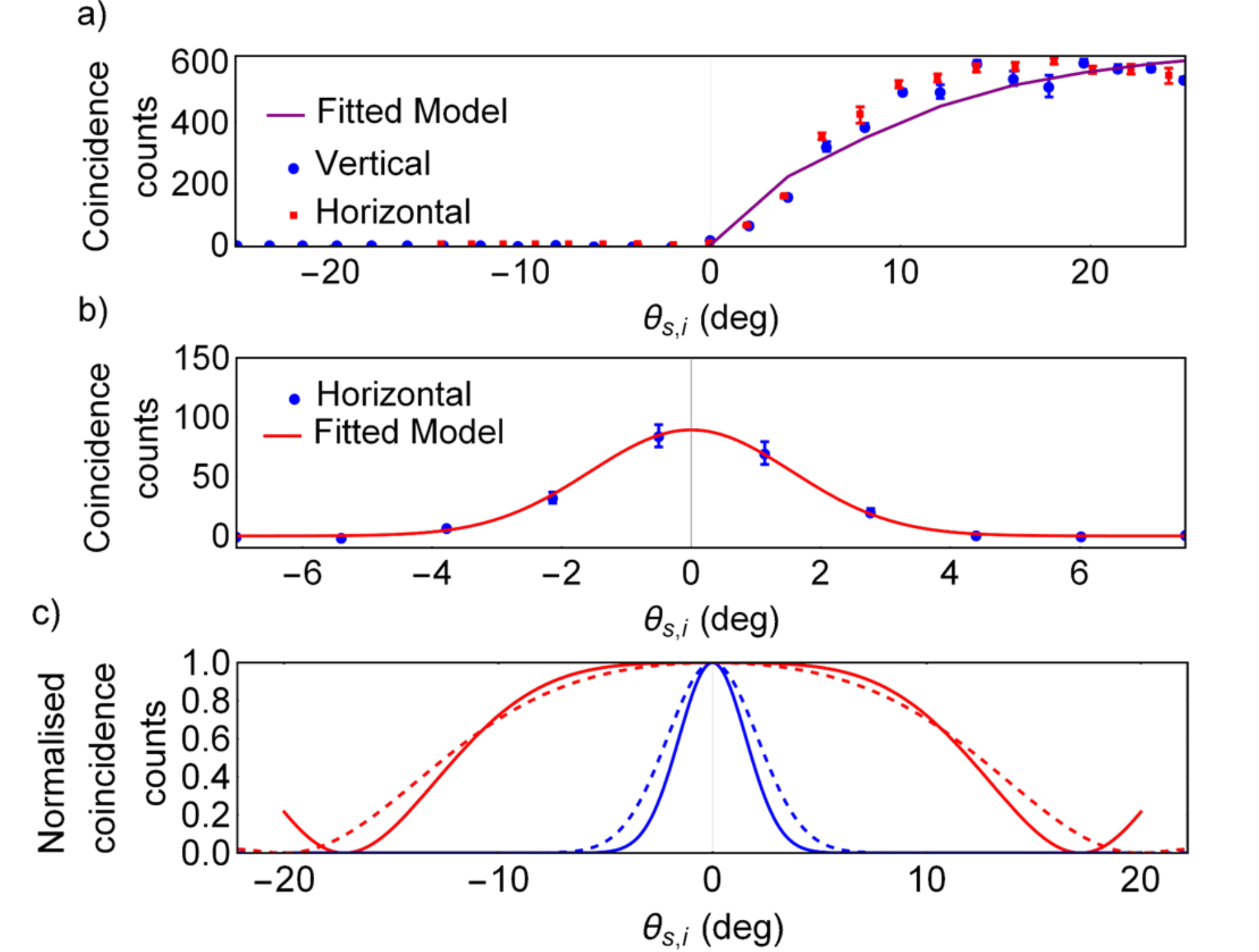} 
\caption{Reconstruction of the unconditional and conditional angular distributions. a) Number of coincidences acquired in $70$ minutes for different positions of the knife edge (points) and their numerical fit (line). b) Number of coincidences acquired in $30$ minutes for different positions of the slit (points) and the expected Gaussian fit (line). The fitted models from a) and b) are plotted in c), where the blue line is the unconditional width and the red is the conditional width, and compared to the theoretical distributions (dashed lines).}
\label{fig:knifeslit}
\end{figure}

To measure the total angular range of SPDC, the radiation was sent through a scanning knife edge. The radiation was then split into two arms, coupled into two mulitmode fibers (MMF) and sent to a Hanbury Brown-Twiss (HBT) setup and the number of coincidence counts was recorded. The photons in each pair were anti-correlated in angle, this meant that scanning a knife edge across the angular range of the emission cut both the positive and negative angular ranges simultaneously. To ensure that we only registered SPDC correlations, the number of accidental coincidences was subtracted from the total number of coincidences. The number of accidental coincidences was found by measuring the total number of coincidences in a time window far from the arrival time difference of the signal and idler photons from LN (see inset in Fig.~\ref{fig:setup1}). Type-0 SPDC emission is azimuthally symmetric, therefore the number of coincidence counts should be given by the angular spectrum of SPDC, integrated from the knife edge position to the position of collinear emission ($\theta_{i,s}=0^{\circ}$). Fig.~\ref{fig:knifeslit}a shows the number of coincidences at different knife edge positions (points), with a numerical fit of the expected distribution~\cite{monken1998transfer} (line). Blue and red points correspond to scanning in the vertical and horizontal directions, respectively. Their overlap confirms the azimuthal symmetry of the experiment.

To measure the conditional width of the TPI, the scanning knife edge was replaced by a slit in front of the collimated emission. Again the emission was split and sent to a HBT setup and the number of coincidences was measured. The slit was scanned across the collinear direction of the radiation. The results are shown in Fig.~\ref{fig:knifeslit}b, where the number of real coincidences is shown in  blue and a Gaussian fit is shown in red. The fitted conditional and unconditional angular distributions are then compared in Fig.~\ref{fig:knifeslit}c (solid lines) with the corresponding theoretical distributions (dashed lines).

\begin{figure}
\includegraphics[width=8cm]{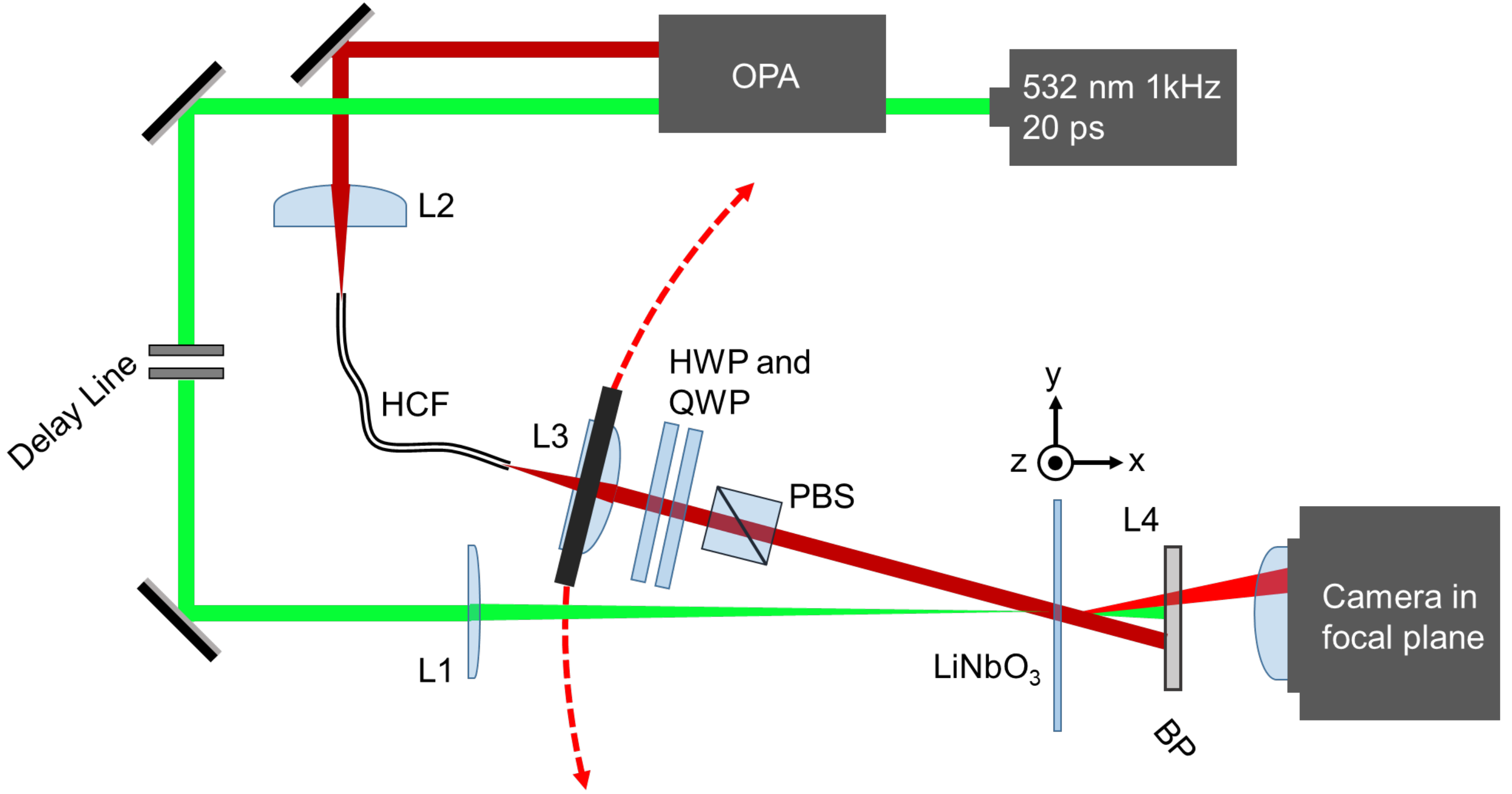} 
\caption{The setup for angular SET. Lens L3 is placed on a rotation platform along with the half-waveplate (HWP), quarter-waveplate (QWP), and polarising beam splitter (PBS). BP is a bandpass filter centered at $800$ nm with a bandwidth of $10$ nm. HCF is a hollow-core fiber. Lenses L3 and L2 are used to couple in and out of the HCF, whilst lens L1 is interchangeable to control the beam waist.}
\label{fig:setsetup}
\end{figure}

\begin{figure}
\includegraphics[width=8.5cm]{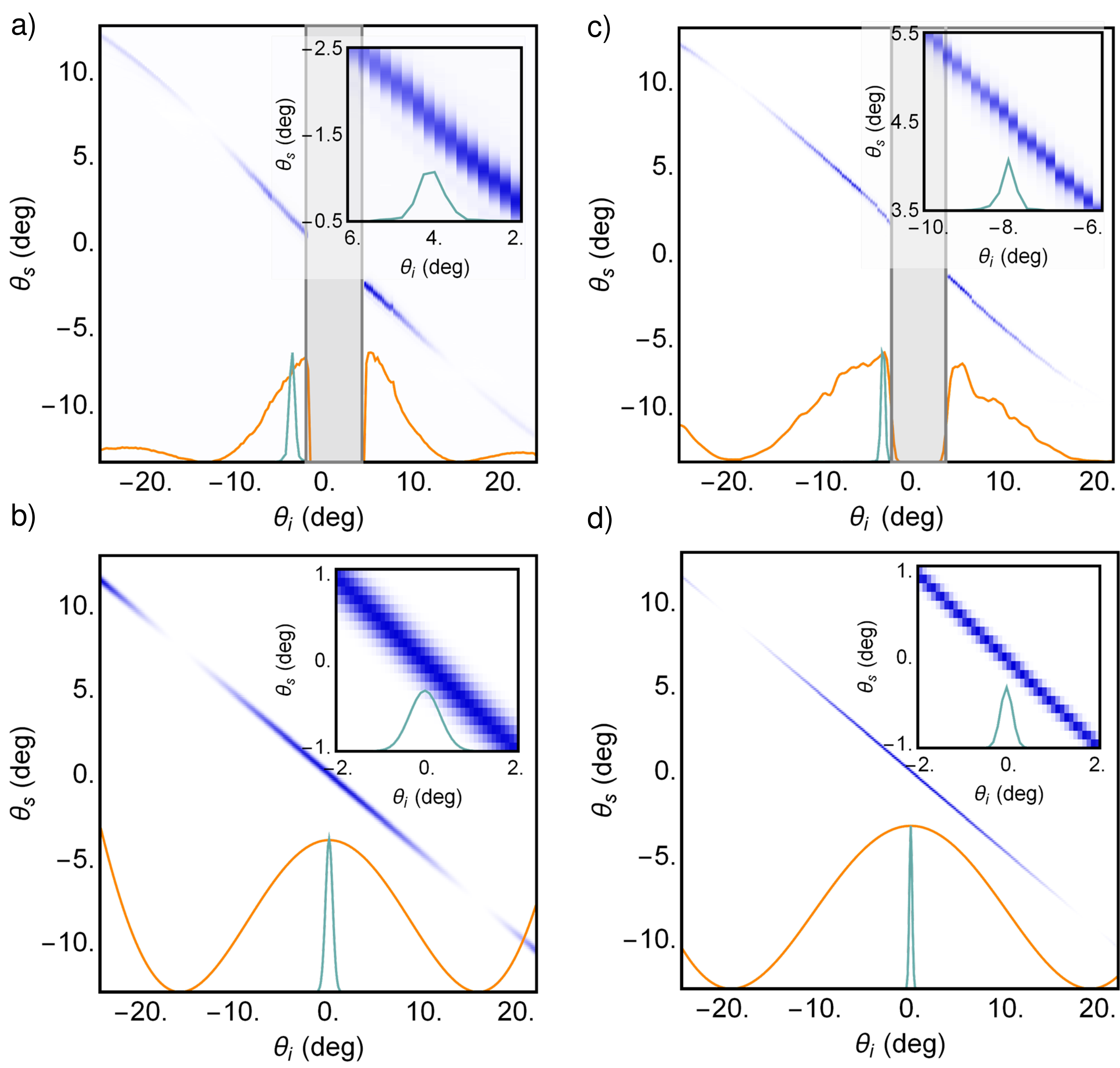} 
\caption{Angular TPI for non-phase matched SPDC measured via SET. Top panels: the TPI measured for a) pump beam waist of $\sigma = 60~ \mu \text{m}$ and interaction length  $L= 6.7~ \mu$m and c) pump beam waist of $\sigma = 120~ \mu \text{m}$ and interaction length  $L= 6.4~ \mu$m. Bottom panels show the simulated TPI for the corresponding cases. The insets show a zoom of the TPI to see the scale of the conditional curve (teal).}
\label{fig:setjai}
\end{figure}

To measure the TPI in the case of softly focused pump, leading to high degree of entanglement, stimulated emission tomography (SET)~\cite{liscidini2013stimulated} was implemented. A seed beam stimulated the emission of the idler photon, which led to enhanced emission in the signal mode. Whilst SET is typically implemented in the frequency domain, here we have used it to probe the TPI in angular space. SET allowed us to measure the signal emission directly using a SPIRICON camera (Fig.~\ref{fig:setsetup}), without resorting to single-photon detectors. To reconstruct the full angular TPI, we stimulated the idler at all possible emission directions. This was done by changing the incident angle of the seed beam impinging on the LN sample. The second harmonic of a Nd:YAG laser at $532$ nm, with a $20$ ps pulse width and $1$ kHz repetition rate, was used as the pump. The seed was centered at $1600$ nm with the same pulse properties as the pump and a pulse energy of 10 $\mu$J. It was coupled into a high damage threshold hollow-core fiber (HCF). The out-coupler was placed on a rotation platform and the seed beam sent through a half-wave plate (HWP), quarter-wave plate (QWP) and polariser (PBS) to ensure the seed was polarised along the z axis of the LN crystal. A delay line ensured that the pump and seed pulses arrived simultaneously at the LN. Lens L1 was used to focus the pump and to check how the degree of entanglement changed with the beam size; it was either chosen to have a focal length of 100 mm or 200 mm. The LN was placed at the focal point of the pump, coinciding with the center of rotation of the platform.
The seed was unfocused on the sample, so that it could be approximately described by a plane wave, compared to the pump beam. To reconstruct the TPI, the angle of incidence of the seed was scanned and the intensity distribution of the signal ($797$ nm) beam was recorded in the Fourier plane (Fig.~\ref{fig:setjai}). To filter out the pump and seed beam a bandpass filter (BP) was placed before the camera. To calibrate the emission angles of the signal beam, the second harmonic generation ($800$ nm) of the seed beam was used. 

The TPI was measured at two positions on the LN, corresponding to $L=6.3\,\mu$m and $L=6.6 \,\mu$m, using two different lenses (L1), with focal lengths $200$ mm and $100$ mm, respectively. Fig.~\ref{fig:setjai}a,c demonstrates the TPI measured for these two sets of parameters. The data between roughly $-3^\circ$ and $3^\circ$ is missing because the optics behind the seed beam blocked the pump. The small discrepancy between the theory (Fig.~\ref{fig:setjai}b,d) and experiment can be attributed to both lens aberrations and reflection losses at large incident and collection angles.   

The ratio between the emission (unconditional) width $\Delta=19.5^{\circ}$ and the correlation (conditional) width $\delta=0.5^{\circ}$, from Fig.~\ref{fig:setjai}b, gives $R_{1D}=39$. This is in good agreement with the theoretical value of $R_{1D}=37$. Accounting for both transverse coordinates, the total Fedorov ratio across a cross section of the beam is roughly $R_{2D}=1200$. The Schmidt number for the theoretical TPI in Fig.~\ref{fig:setjai} a gives $K=39$, which agrees well with the Fedorov ratio. Note that both these numbers underestimate the degree of entanglement as they do not take into account the high side lobes of the TPI~\cite{reichert2017quality}.

In conclusion, we have shown that photon pairs emitted from an ultrathin layer of lithium niobate via non-phase matched SPDC display huge transverse momentum entanglement at a fixed frequency. The state generated in such a process is not dissimilar to the state imagined by Einstein, Podolsky and Rosen in 1935 and by moving to even thinner platforms on which to generate photon pairs it may be possible to closely imitate the system they proposed. Although the rate of two-photon emission from non-phase matched SPDC is low, there is still scope to optimise this. For example, semiconductors promise far higher second-order susceptibilities, which could dramatically improve the emission rate. In addition, structuring materials may lead to a Purcell enhancement~\cite{davoyan2018quantum}. With higher two-photon emission rates, and the high degree of entanglement reported here, non-phasematched SPDC could surpass any source reported so far.

Not only does the large degree of entanglement mean a large quantum information capacity, it can also improve the resolution of imaging with quantum light. Several quantum imaging techniques have been proposed and implemented over the years, most of them based on entangled photons~\cite{brida2010experimental, lemos2014quantum, pittman1995optical}. The spatial resolution of many of these techniques, as with all imaging techniques, is limited by the range of transverse wavevectors emitted. Due to the large number of transverse wavevector modes generated from non-phase matched thin layers, using SPDC generated from a thin layer should improve the resolution limits of quantum imaging techniques. 

Here we have only investigated the far-field correlation distribution experimentally, an interesting step would be to investigate the near-field correlations (Fig. \ref{fig:theory}b). In the near field the crystal length determines the correlation width in position. Fundamentally, the interaction length is limited by an atomically thick monolayer, such as molybdenum disulfide \cite{saleh2018towards}. Using a monolayer would increase the correlation resolution in position to the deeply subwavelength regime.  Implementing non-phase matched SPDC as a pump in a two photon microscopy setup and relying on the excitation of a fluorescent (or absorbent) material by correlated photons would allowing imaging well beyond the Abbe diffraction limit.

\bibliographystyle{ieeetr}
\bibliography{bibliography}


\newpage

\section{\label{sec:Supplementary} Supplementary information}

{\it Knife edge reconstruction- } For an interaction length limited by the non-linear material boundaries and interaction area limited by the pump beam waist the form of the pump and phase matching functions is
\begin{eqnarray}
    F_{p} (\Delta k_{\bot})=\exp\left(\frac{\Delta k_{\bot} \sigma^2}{2}\right),\\
    F_{pm} (\Delta k_{\parallel})=\sinc \left(\frac{\Delta k_{\parallel} L}{2} \right)^2
\end{eqnarray}
where the pump beam waist $\sigma$ is the width of one standard deviation. The fit for the slit measurement was simply given by $F_{p}$. As this is a Gaussian function there is no difference in width between a 1D distribution and a 2D symmetric distribution. To fit the knife edge measurement the integrated value of of $F_{pm}$ was taken in two dimensions, $\Delta k_{\parallel}=\sqrt{\Delta k_{x}^2+\Delta k_{y}^2}$. To account for the scanning knife edge and because we were collecting coincidences, the integration boundaries over $k_{x}$ were reduced symmetrically. The argument $\Delta k_{\parallel}(\theta_i,\theta_s)$ was used as the fitting parameter and the integrated sinc function was fitted by hand to match the curve in Fig~\ref{fig:knifeslit} b.

{\it Collection efficiency in the spontaneous regime- } Non-phase matched SPDC is not only inefficient but also difficult to collect due to the highly multimode nature of the emission. Although we were working with multimode detectors, spherical aberrations and differing mode divergences led to a restricted number of modes able to be detected efficiently. Reducing the number of modes by reducing the pump beam waist led to an increase in detection efficiency at the expense of low entanglement. This is the reason for choosing a high NA lens for L3 in the correlation experiment. Increasing the efficiency of the non-phase matched source would be one way to increase the detection probability of a highly entangled state. Similarly, improving detection losses in the setup would allow us to detect a more highly entangled state. The quantum efficiency of the mulitmode detectors was around 50 \% leading to a correlation efficiency of 25 \%. The bandpass filter (IF) used had a transmission of 50 \% and bandwidth of 10 nm. Internal reflection caused by the high refractive index of the LN dropped the efficiency by an additional 20 \%. These values could be improved to yield higher correlation rates, for example by optimising the bandpass filter, using anti-reflection coating at the correct frequency and using superconducting nanowires as opposed to avalanche diodes. 
Lastly, the huge angle of emission of non-phase matched SPDC requires a large NA lens to capture it. For the correlation measurement the lens (L4) had an NA = 0.3. This coincided almost with the emission angle of the radiation expected theoretically. Without knowing the crystal length precisely at a given point on the LN wafer, it was difficult to discern whether the lens aperture limited the emission angle.



\end{document}